%% file: ms2622.tex
\documentclass[referee]{raa}
\usepackage{graphicx,times}
\usepackage{natbib}
\usepackage{amssymb,amsmath}

\usepackage[a4paper=true,dvipdfm=true,pagebackref=true]{hyperref}
\hypersetup{pdftitle = The title of my PDF, pdfauthor = My name, pdfsubject= The subject, pdfkeywords = keyword1 keyword2 keyword3} 
\hypersetup{colorlinks = true, linkcolor = green, anchorcolor = red, citecolor = blue, filecolor = red, pagecolor = red, urlcolor = red}

\begin{document}

    \title{FGK 22 $\mu$m Excess Stars in LAMOST DR2 Stellar Catalog}

 \volnopage{ {\bf 2012} Vol.\ {\bf X} No. {\bf XX}, 000--000}
   \setcounter{page}{1}

\author{Chao-Jian Wu\inst{1,2}, Hong Wu\inst{1,2}, 
Kang Liu, \inst{3}, Tan-Da Li\inst{2}, 
Ming Yang\inst{1}, Man I Lam \inst{1}, 
Fan Yang\inst{1}, Yue Wu\inst{1}, 
Yong Zhang\inst{4}, Yonghui Hou\inst{4}, 
Guangwei Li\inst{1}}

\institute{Key Laboratory of Optical Astronomy, 
National Astronomical Observatories, Chinese Academy of Sciences，
Beijing 100012, China; {\it chjwu@bao.ac.cn}\\
\and
National Astronomical Observatories, 
Chinese Academy of Sciences, Beijing 100012, China\\
\and
Department of Astronomy, Beijing Normal University,
Beijing, 100875, P.\ R.\ China\\
\and
Nanjing Institute of Astronomical Optics \& Technology, 
National Astronomical Observatories, Chinese Academy of Sciences, 
Nanjing 210042, China\\
\vs \no
{\small Received ---; accepted ---}
}

\abstract{
    Since the release of LAMOST (the Large Sky Area Multi-Object Fiber 
    Spectroscopic Telescope) catalog, we have the opportunity to use 
    the LAMOST DR2 stellar catalog and \emph{WISE All-sky catalog} to 
    search for 22 $\mu$m excess candidates. In this paper, we present 10 
    FGK candidates which show an 
    excess in the infrared (IR) at 22 $\mu$m. The ten sources are all 
    the newly identified 
    22 $\mu$m excess candidates. Of these 10 stars, 5 stars are F type 
    and 5 stars are G type. The criterion for selecting candidates is 
    $K_s-[22]_{\mu m}\geq0.387$. In addition, we present the spectral 
    energy distributions (SEDs) covering wavelength from optical to 
    mid-infrared band. Most of them 
    show an obvious excess from 12 $\mu$m band and three candidates even 
    show excess from 3.4 $\mu$m. To characterize the amount of 
    dust, we also estimate the fractional luminosity of ten 22 $\mu$m 
    excess candidates.  
}
\authorrunning{Ch.-J. Wu et al.}
\titlerunning{22 $\mu$m Excess Stars in LAMOST DR2}
\maketitle

\section{Introduction}
\label{s:intro}
More and more surveys (\emph{IRAS}, \emph{Infrared Space Observatory (ISO)},
\emph{Spitzer Space Telescope}, \emph{Herschel}, 
\emph{WISE}, etc.) have 
been conducted to search for IR excess stars (especially the dusty 
disks and exoplanets system) since the first discovery of the IR 
excess phenomenon (known as a debris disk) around the Vega in 1980s 
\citep{aumann84}.

In fact, there are many reasons that can cause excess at IR band 
\citep{wu2013}. Protostars \citep{thompson1982}, surrounding dust 
disk \citep{gorlova2004, gorlova2006, rhee2007, hovhannisyan2009,
  koerner2010, wu2012}, companion star (like white dwarfs + M 
stars, white dwarfs + brown dwarfs or + dusty disks) \citep{debes2011}, 
giant stars, background galaxy, background nebula, interstellar medium 
and random foreground object \citep{ribas2012} could all produce 
infrared excess. 
In our study, we only focus on the IR excess stars 
which are caused by protostars and surrounding dust disk.

In previous work, there is not a complete stars catalog 
with so much spectral information to be used for searching IR excess stars. 
The LAMOST \citep{cui2012} Data Release 2 (DR2) contains 4,136,482 spectra 
and it is the largest stellar spectral catalog in the world 
at present. Undoubtedly, it will provide us with unique insights into 
the population of IR excess stars. More detailed about LAMOST data will 
be described in Sec \ref{s:data:lamost}. 

While searching the infrared excess stars from LAMOST Stellar Catalog, 
we also introduce the Wide-field Infrared Survey Explorer (WISE; 
\citeauthor{wright2010}~\citeyear{wright2010}) All-sky data 
\citep{wu2013}. 
Some recent work about studying IR excess stars with \emph{WISE} 
have been published. \citet{rizzuto2012} presented 
an analysis of 829 \emph{WISE} stars in the Sco-Cen OB2 association 
and observed that B-type stars have a smaller excess fraction than 
A and F-type stars. \citet{luhman2012} found $\sim50$ new transitional, 
evolved and debris disks candidates from \emph{Spitzer} and \emph{WISE} 
data. \citet{wu2013} focused on the bright stars ($V_{mag}\leq10.27$) and 
searched more than 70 new identified 22 $\mu$m excess stars. 
\citet{vican2014} studied 2820 solar type stars and found 74 new 
stars with \emph{WISE} excess at 22 $\mu$m. \citet{patel2014} 
presented a sensitive search for 
\emph{WISE} W3 and W4 excess candidates within 75 pc from the Sun 
and expanded the number of known 10-30 $\mu$m excesses to 379. 
\citet{theissen2014a, theissen2014b, theissen2014c} studied M 
dwarfs with \emph{WISE} 12 and 22 $\mu$m excess. 

In this paper, we introduce LAMOST Survey and describe the candidate 
selection criterion, source identification in Section \ref{s:data}. 
In Section \ref{s:results}, we analyze their IR properties, and 
present their spectral energy distributions (SEDs). The conclusion 
and summary are presented in Section \ref{s:summary}.

\section{Data}
\label{s:data}

\subsection{LAMOST Survey}
\label{s:data:lamost}

LAMOST, also known as Wang-Su Reflecting Schmidt 
Telescope, is a new type of wide field telescope with a large aperture. 
The advantages of wide field and large aperture, combined with 4000 
fibers (taking 4000 spectra in a single exposure), provide us with 
opportunity to carry out even the largest stellar and galactic 
survey \citep{zhao2012}.
The limiting magnitude is as faint as r=19 at the resolution R=1800, 
which is equivalent to the design aim of r=20 for the resolution R=500. 

To maximize the scientific potential of the facility, wide national 
participation 
and international collaboration have been emphasized. The survey 
has two major components: the LAMOST ExtraGAlactic Survey (LEGAS) 
and the LAMOST Experiment for Galactic Understanding and 
Exploration (LEGUE) survey of Milky Way stellar structure.

The first observation mission of the LAMOST regular survey launched 
on September 28, 2012 \citep{luo2012}, and have been already 
successfully finishied on July 15, 2013. Totally, over 1.2 million 
spectra with signal to noise larger than 10 of 689 square degrees 
are obtained. The data set of LAMOST data release one (DR1), 
including spectra of the pilot survey and spectra of the first 
year of the regular spectroscopic 
survey, have already published to domestic data users and foreign 
partners. The DR1 totally contains 2,204,860 spectra, including 
717,660 spectra of pilot survey and 1,487,200 spectra of regular 
survey. 

In the stellar catalog, there are 648,820 stars in pilot 
survey and 1,295,586 stars in the first year spectroscopic survey. 
In addition, the stellar catalog contains the atmospheric 
parameters of 1,085,404 stars, which becomes the largest stellar 
spectral parameters catalog in the world at present.

The secend year observation mission of the LAMOST regular survey 
launched on September, 2013, and have been already successfully 
accomplished on July ,2014 under the joint effort 
of entire staffs 
of the center for operation and development of Guoshoujing telescope. 
Totally, over 1.3 million spectra with signal to noise larger than 
10 are obtained during the past year, which 
sufficiently demonstrate the advantage of LAMOST in spectra accessing. 
The data set of LAMOST data release two (DR2), including spectra of 
the pilot survey and spectra of the past two years of the regular 
spectroscopic survey, have already published to domestic data users 
and foreign partners. The DR2 totally contains 4,136,482 spectra, 
including 909,520 spectra of pilot survey and 3,226,962 spectra of 
regular survey. In addition, the DR2 contains the atmospheric parameters 
of 2,207,788 stars, which becomes the largest stellar spectral 
parameters catalog in the world at present.
\footnote{www.lamost.org}

\subsection{Data Selection}
\label{s:data:sel}

As described above, we mainly used the \emph{WISE All-sky} catalog 
and LAMOST DR2 stellar catalog to search IR excess stars in this paper. 
Before doing the cross-match, we chose the FGK stars  
from LAMOST DR2 stellar catalog, because the FGK stars have more 
reliable photometric information. Then we uploaded the coordinates of 
chosen FGK stars into WISE all-sky catalog download website 
with matching radius $6"$ \citep{wu2013} and $cc_{flags}=0000$ 
(Contamination and confusion flag), 5,511 stars are obtained. 

The Signal to Noise ($S/N$) is an important index for photometric 
precision and it affects the selecting accuracy of candidates 
directly. So we deleted those sources with $S/N\le15$ in the 
W1, W2, W3, W4, $g$ and $i$ band from 5,511 sources. Then 2,214 
sources are left.

To reduce the effection of variable star, 
contamination and bad pixels in the CCD, we considered the 
related index ($var\_flg$ and $w?flg$ Symbol ? 
indicates w1, w2, w3 and w4 band) provided in the WISE 
\emph{All-sky} catalog. These indexes can help us select out 
stars with much higher photometric accuracy. To this step, 2,134 
sources are obtained.

\begin{figure*}                                                               
    \includegraphics[width=\textwidth]{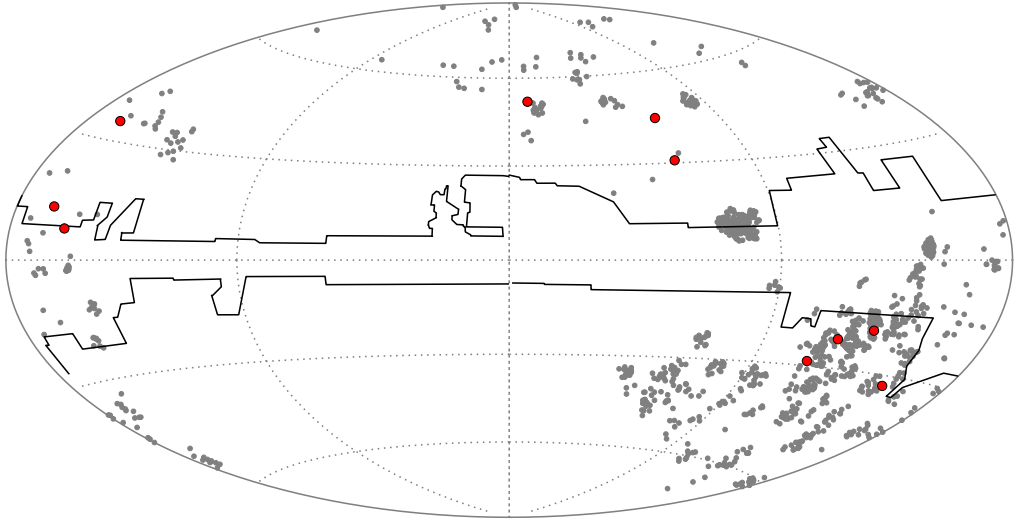}
    \caption{Distribution of tne 22 $\mu$m excess stars in a Galactic 
    Aitoff projection. The red points are the 10 IR-excess candidates 
    and the gray points are the distribution of the matched sources 
    from LAMOST DR2 stellar catalog. The molecular cloud and the 
    star formation region locate between the two black solid lines.
    \label{fig:allsky_dist}}
\end{figure*}

Moreover, there is another important index, 
saturation! In the WISE catalog, saturation begins to occur for 
point sources brighter than $w1\sim8$, $w2\sim7$, $w3\sim4$ and 
$w4\sim0$ mag. By this criterion, there will be no saturated 
stars in our candidates.
Last, we obtained 1,120 stars (gray regions in Figure
\ref{fig:allsky_dist}.) 

\begin{figure*}                                                              
    \includegraphics[width=\textwidth]{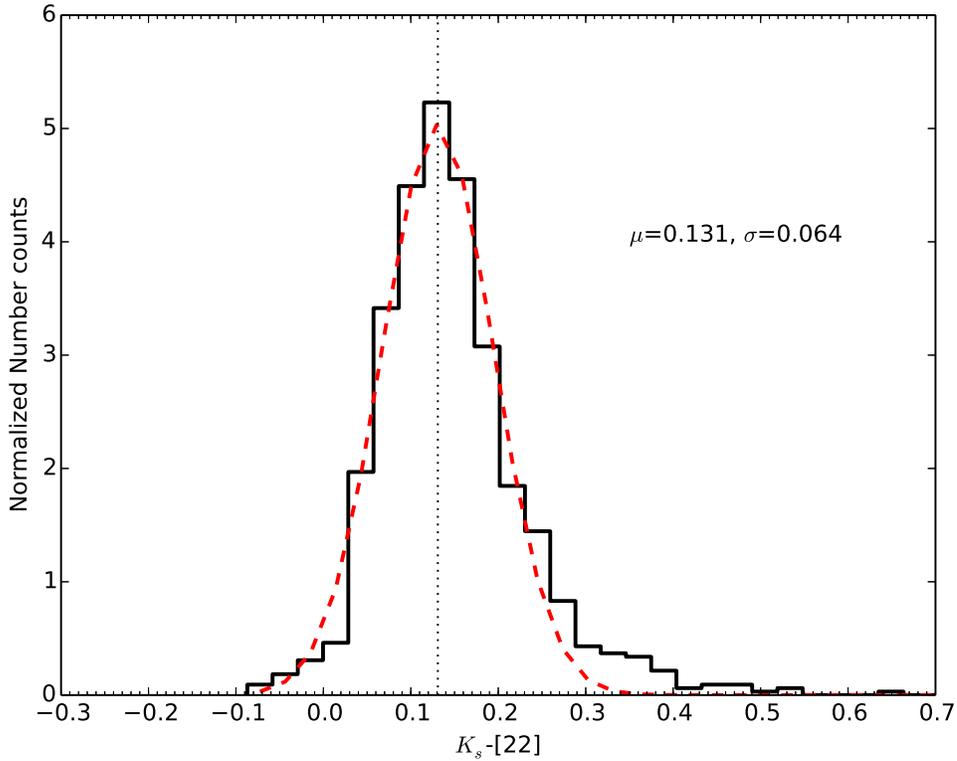}
    \caption{Goodness of fit for all matched sources. The criterion is 
    $K_s-[22]_{\mu m}\geq \mu+4\sigma = 0.387$, which means that those with 
    $K_s-[22]_{\mu m}\geq0.387$ can be identified as 22 $\mu m$ excess stars.
    \label{fig:hist_k22}}
\end{figure*}

\begin{figure*}                                                              
    \includegraphics[width=\textwidth]{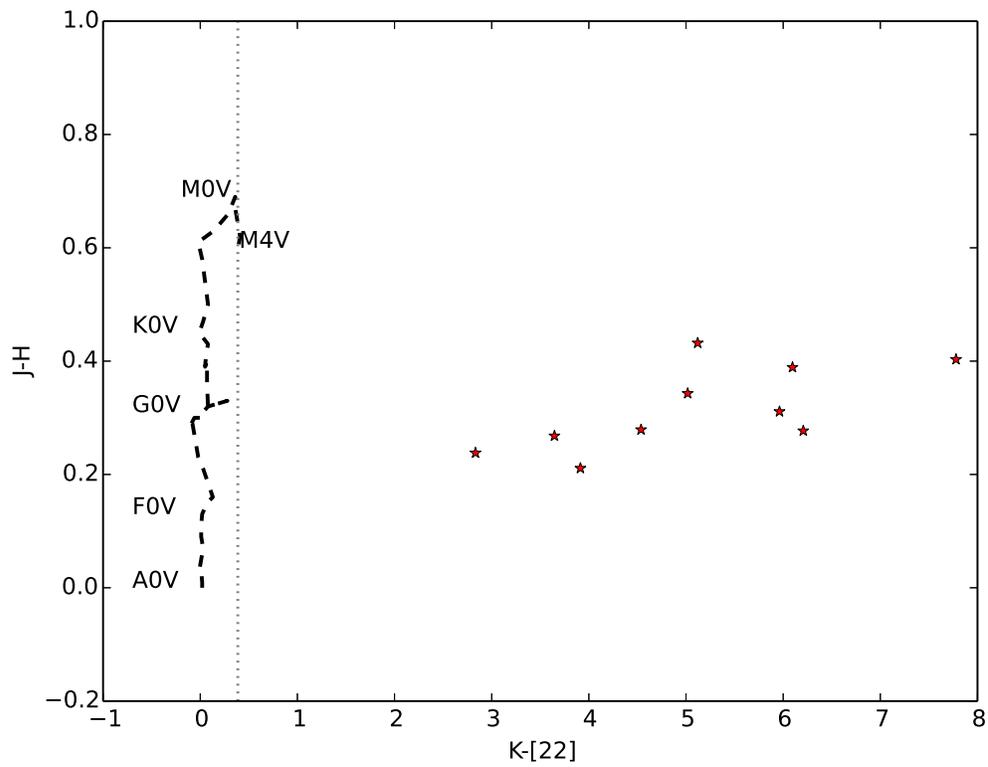}
    \caption{Diagram of $J-H$ vs. $Ks-[22]$. The red star symbols 
    show the distribution of FGK stars. Normal dwarf stars are 
    plotted as black dashed line, the corresponding spectral types 
    are also labeled. The gray dotted line gives the criterion 
    of the 22$\mu$m excess candidates.
    \label{fig:k22jh}}
\end{figure*}

The method of selecting 22 $\mu$m excess stars is similar with our 
previous work \citep{wu2013}. 
In this work, we  gave the $\mathrm{K_s}-[22]$ criterion by 
fitting the histgram of $\mathrm{K_s}-[22]$ with a gaussian 
function. Figure \ref{fig:hist_k22} shows the histgram  
of $\mathrm{K_s}-[22]$. To reduce the bias introduced by the 
small number of matched catalog, we use 4$\sigma$ confidence for all the 
1,120 sources. That means the criterion used in this work is 
$\mathrm{K_s}-[22]\geq\mu+4\sigma=0.387$. The $J-H$ vs. $K_s-[22]$ 
diagram can also 
help to reject those with incorrect spectral type by comparing with 
normal dwarf stars (dashed line in Figure \ref{fig:k22jh}). 
There are 64 sources with $\mathrm{K_s}-[22]\geq0.387$. 
To remove the background contamination, we selected those with 
\emph{IRAS} 100 $\mu$m backgound level lower than 5 MJy$sr^{-1}$ 
\citep{kennedy2012, wu2013}.
Then 54 sources are eliminated and ten [22] $\mu$m excess candidates 
are obtained lastly.
(red points in Figure \ref{fig:allsky_dist}).

\section{Analysis and Results}
\label{s:results}

\subsection{Notes on candidates}
\begin{table}
\bc
\begin{minipage}[]{100mm}
\caption[]{10 mid-infrared excess stars from LAMOST DR2 Stellar Catalog
\label{tab:catalog}}
\end{minipage}
\setlength{\tabcolsep}{1pt}
\small
\begin{tabular}{lcccccccccccc}
\hline\noalign{\smallskip}
ID & RA & DEC & K-[22] & log $g$ & e\_log $g$ & $T_{eff}$ & e\_$T_{eff}$ & [Fe/H] & e\_[Fe/H] & $R_v$ & SPT & $f_d$ \\
(LAMOST) & (J2000) & (J2000) & (mag) & & & (K) &  & & & & &\\
    \hline\noalign{\smallskip}
   J000153.53+330250.0 & 0.4730602   & 33.0472467 & 5.017 & 4.527 & 0.518 & 5666.7 & 224.52 & 0.004 & 0.242 & 14.02  & G6 & 2.3e-3  \\
   J002623.58+403943.1 & 6.5982554   & 40.6619950 & 7.779 & 4.117 & 0.622 & 5905.99& 222.03 & -0.228 & 0.237 & -58.37 & G2 & 3.5e-2  \\
   J012825.91+432504.2 & 22.1079984  & 43.4178478 & 3.913 & 4.209 & 0.485 & 6242.06& 159.5  & -0.146 & 0.165 & -39.14 & F6 & 4.2e-4  \\
   J023430.10+243831.9 & 38.625456   & 24.6422148 & 5.120 & 4.713 & 0.331 & 5068.22& 83.28  & -0.006 & 0.116 & -3.16  & G9 & 2.1e-3  \\
   J070240.89+114104.0 & 105.6703877 & 11.6844692 & 6.207 & 4.379 & 0.481 & 6199.84& 176.02 & -0.011 & 0.175 & -13.49 & F2 & 9.0e-3  \\
   J070904.25+202205.4 & 107.2677215 & 20.3681877 & 6.096 & 3.722 & 0.739 & 5530.51& 183.66 & -0.596 & 0.233 & 46.47 & G3 & 8.9e-3  \\
   J083820.18+283823.0 & 129.5841041 & 28.6397304 & 2.833 & 4.343 & 0.420 & 6201.13& 86.15  & -0.175 & 0.101 & 31.7   & F7 & 2.1e-4  \\
   J151031.26+072454.1 & 227.6302529 & 7.415044   & 3.646 & 4.301 & 0.470 & 6034.04& 110.4  & -0.4   & 0.136 & 11.88  & F7 & 3.9e-4  \\
   J162412.18+385720.8 & 246.0507688 & 38.9558047 & 5.962 & 4.33  & 0.480 & 6255.22& 206.16 & -0.364 & 0.237 & -61.45 & F5 & 3.5e-3  \\
   J173214.60+360622.1 & 263.0608529 & 36.106158  & 4.537 & 4.144 & 0.548 & 5914.00& 143.17 & 0.159  & 0.136 & -49.44 & G3 & 1.1e-3  \\
    \noalign{\smallskip}\hline
    \end{tabular}
    \ec
    Column 1: Names of candidates\\
    Column 2-3: Coordinates of candidates\\
    Column 4: K-[22] - Criterion of selecting 22 $\mu$m excess\\
    Column 5: log$g$ - Surface Gravity\\
    Column 6: e\_log$g$ - Surface Gravity Uncertainty\\
    Column 7: $T_{eff}$ - Effective Temperature\\
    Column 8: e\_$T_{eff}$ - Effective Temperature Uncertainty\\
    Column 9: [Fe/H] - Metallicity\\
    Column 10: e\_[Fe/H] - Metallicity Uncertainty\\
    Column 11: $R_v$ - Heliocentric Radial Velocity\\
    Column 12: SPT - Spectral type\\
    Column 13: $f_d$ - Fractional Luminosity
\end{table}

Table \ref{tab:catalog} shows the selected 10 IR excess candidates.
The names of the 10 candidates are from \emph{WISE} catalog. 
From the values of $K-[22]$ list in Table \ref{tab:catalog} we can 
see that all the ten candidates show obviously excess in the 22 
$\mu$m. log$g$, $T_{eff}$, [Fe/H], $R_v$ and Spectral type are the 
stellar parameters provided by LAMOST DR2. The fractional luminosity 
$f_d$, which is defined as the ratio of integrated infrared excess 
of the disk to bolometric luminosity of star, is also estimated for 
each candidate. More detailed can be seen in Section \ref{s:fd}.

Compared with database (like \emph{SIMBAD}) and previous work, they 
are all the newly identified 22 $\mu$m excess candidates. The spectral 
type of the ten candidates range from F2 to G9. 

\subsection{Hertzsprung-Russell Diagram}
\begin{figure*}                                                              
    \includegraphics[width=\textwidth]{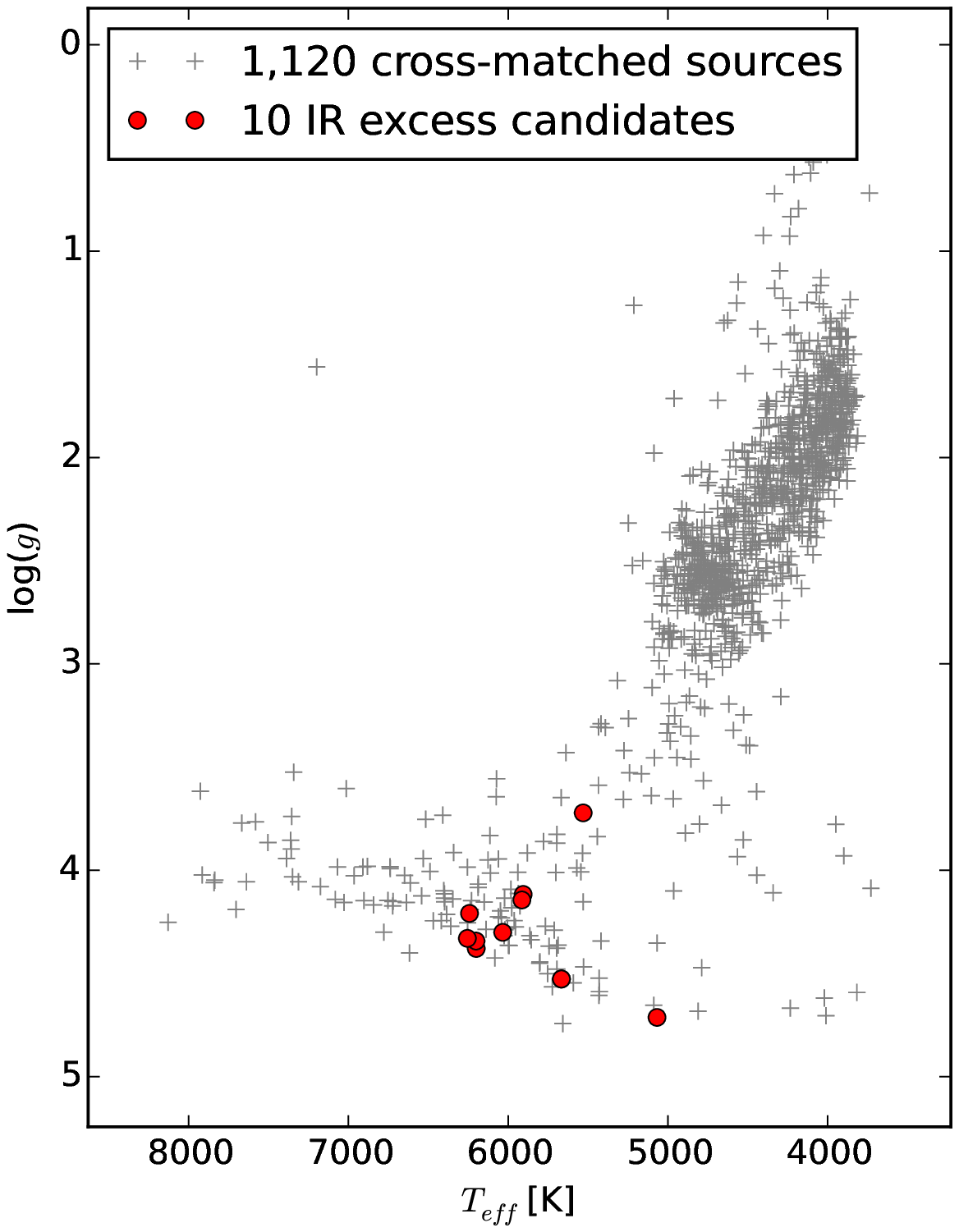}
    \caption{H-R Diagram of 22 $\mu$m execss stars. Gray +  symbols 
    are the more than selected 1000 sources from LAMOST DR2 catalog. 
    Red points are the 10 candidates with excess IR, 9 of them are 
    obviously the main-sequence stars.
    \label{fig:hrd}}
\end{figure*}

Generally, researches of debris disks or IR excess stars mainly focus 
on the main-sequence stars. However, there are still some work about
giant stars with IR excess (\citep{jones2008, groenewegen2012}). 
The stellar parameters provided in LAMOST DR2 can help us to separate 
main-sequence stars and giants. Figure \ref{fig:hrd} shows the 
temperature $T_{eff}$ versus $log$(g) Hertzsprung-Russell (short for 
H-R) diagram.

From Figure \ref{fig:hrd} we can see that 9 sources are obviously 
main sequence stars. Among the 10 candidates,  5 stars are F type stars 
and 5 G type stars. They are all belong to the $Sunlike$ stars and 
noted in Table \ref{tab:catalog}. 

\subsection{Spectral Energy Distribution}
\label{s:sed}
Figure \ref{fig:sed} shows the SEDs of 10 selected IR excess 
candidates. 
All these SEDs are fitted by 
using optical band 
information provided by \emph{LAMOST} DR2, near infrared band 
(J, H, K, W1, W2, W3 and W4) provided by \emph{2MASS} and 
\emph{WISE}. The fitted stellar parameters are consistent with 
those given by LAMOST. (Table \ref{tab:catalog})
\citep{Robitaille2007,Kurucz1979}. 
All the candidates show an obvious excess at 22 
$\mu$m (W4), even at 12 $\mu$m band (Figure \ref{fig:sed}). 
The up triangular symbols 
mean flux excess comparing to black body radiation. Some even 
show excess in the 3.4 $\mu$m band (\textbf{J002623.58+403943.1}, 
\textbf{J070240.89+114104.0} and \textbf{J070904.25+202205.4}), 
which indicates that they are 
surrounded by hotter dust than other candidates.

Since there is no photometric information in longer band, we 
do not know the peak of the infrared excess and can not determine 
the shape of disks around candidates. Therefore, we need 
far-infrared observation to confirm in the future.

\include{ms2622fig5}

\subsection{Fractional Luminosity}
\label{s:fd}

The fractional luminosity $f_d$ ($f_d=L_{IR}/L_*=F_{IR}/F_*$) can 
be used to characterize the amount of dust. So we estimated $f_d$ 
in this paper. We assumed $\nu L_{\nu}$ as the total IR luminosity 
$L_{IR}$ because of the absence of longer bands in WISE. Detailed 
method can be seen in \citet{wu2013}. 

The fractional luminosities of ten candidates are listed in Table 
\ref{tab:catalog}. All the $f_d$s mainly range from $10^{-4}$ to 
$10^{-3}$. Only the $f_d$ of \textbf{J002623.58+403943.1} is even 
lager than $10^{-2}$.  It has been argued that debris disks are 
confined to $f_d < 10^{−2}$. So \textbf{J002623.58+403943.1} probably 
contains a significant amount of gas \citep{artymowicz1996,wu2012}. It is 
worth to verified in the future work. There is no consensus on the 
relation between $f_d$ and ages \citep{wu2013}. So we can not conclude whether 
these candidates are young or old.

\section{Summary}
\label{s:summary}
In this work, we use LAMOST DR2 stellar catalog and \emph{WISE 
All-sky catalog} to search for 22 $\mu$m excess candidates. 
The searching method used in this paper is $J-H$ vs $K_s-[22]$ 
diagram \citep{wu2013}. Then we obtain 10 high-precision candidates  
with 22 $\mu$m excess. 
Each candidate presented here can be studied further with higher 
angular resolutioninfrared imaging or infrared spectroscopy. 
Among the 10 candidates, 5 stars are F type stars and 5 are 
G stars. We also provide the SEDs for all the candidates covering 
wavelength from optical to mid-infrared band. From the SEDs, we 
find that almost all of the 10 candidates show obvious excess in 
12 $\mu$m. There are 3 candidates even show  excess in 3.4 $\mu$m. 
Finally, we estimated the fractional luminosity $f_d$ for each 
candidate. There is one candidate even has $f_d$ larger than 
$10^{-2}$.

\normalem
\begin{acknowledgements}
Ch.-J. Wu thanks C. Liu and Y.-F. Huang for valuable discussion and 
L. Lan for warmhearterd help.
This project is supported by the National Natural Science 
Foundation of China (grant No. 11403061), the China Ministry 
of Science and Technology under the State Key Development 
Program for Basic Research (2014CB845705, 2012CB821800), the 
National Natural Science Foundation of China (grant Nos. 11173030, 
11225316, 11078017, 11303038, 10833006, 10978014, and 10773014), 
the Key Laboratory of Optical Astronomy, National Astronomical 
Observatories, Chinese Academy of Sciences.

Guoshoujing Telescope (the Large Sky Area Multi-Object Fiber Spectroscopic 
Telescope LAMOST) is a National Major Scientific Project built by the 
Chinese Academy of Sciences. Funding for the project has been provided 
by the National Development and Reform Commission. LAMOST is operated 
and managed by the National Astronomical Observatories, Chinese Academy 
of Sciences.

\end{acknowledgements}

\bibliographystyle{raa}
\bibliography{ms2622.bib}

\end{document}

%% file: ms2622fig5.tex
\begin{figure*}[h!tb]
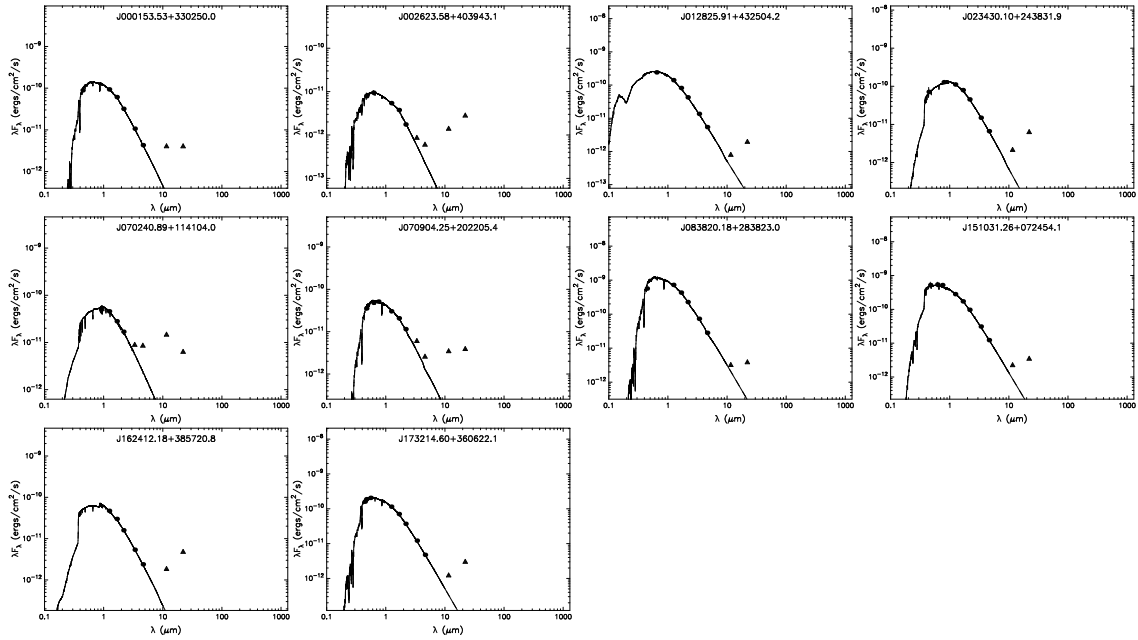

\includegraphics[width=0.25\textwidth]{ms2622fig5a.eps}%
\includegraphics[width=0.25\textwidth]{ms2622fig5b.eps}%
\includegraphics[width=0.25\textwidth]{ms2622fig5c.eps}%
\includegraphics[width=0.25\textwidth]{ms2622fig5d.eps}
\includegraphics[width=0.25\textwidth]{ms2622fig5e.eps}%
\includegraphics[width=0.25\textwidth]{ms2622fig5f.eps}%
\includegraphics[width=0.25\textwidth]{ms2622fig5g.eps}%
\includegraphics[width=0.25\textwidth]{ms2622fig5h.eps}
\includegraphics[width=0.25\textwidth]{ms2622fig5i.eps}%
\includegraphics[width=0.25\textwidth]{ms2622fig5j.eps}%
\caption{The SEDs of 10 candidates. They all almost show excess from 12 $\mu$m 
band. There are 3 candidates even show obvious excess from 3.4 $\mu$m band.
\label{fig:sed}}
\end{figure*}